\documentclass{article}
\usepackage{spconf,amsmath,graphicx,hyperref}
\usepackage{booktabs}
\usepackage{multirow}
\usepackage{cite}
\usepackage{etoolbox} 
\usepackage{setspace} 
\usepackage{enumitem}
\usepackage{xcolor}

\title{Nethira: A Heterogeneity-aware Hierarchical Pre-trained Model for Network Traffic Classification\thanks{$^{\star}$ Corresponding authors.}}
\name{Chungang Lin$^{\dagger\ddagger}$ \qquad Weiyao Zhang$^{\dagger}$ \qquad Haitong Luo$^{\dagger\ddagger}$  \qquad Xuying Meng$^{\dagger\star}$ \qquad Yujun Zhang$^{\dagger\ddagger\star}$}
  
\address{$^{\dagger}$ Institute of Computing Technology, Chinese Academy of Sciences, Beijing, China \\
      $^{\ddagger}$University of Chinese Academy of Sciences, Beijing, China. 
      }

\AtBeginEnvironment{thebibliography}{%
  \fontsize{6pt}{0pt}\selectfont
  \setstretch{0.85}
  \setlength{\itemsep}{0pt}
  \setlength{\parskip}{0pt}%
  \setlength{\parsep}{0pt}%
}

\begin{document}
\maketitle

\begin{abstract}

Network traffic classification is vital for network security and management. The pre-training technology has shown promise by learning general traffic representations from raw byte sequences, thereby reducing reliance on labeled data.
However,
existing pre-trained models struggle with the gap between traffic heterogeneity (i.e., hierarchical traffic structures) and input homogeneity (i.e., flattened byte sequences).
To address this gap, we propose Nethira, a heterogeneity-aware pre-trained model based on hierarchical reconstruction and augmentation.
In pre-training, Nethira introduces hierarchical reconstruction at multiple levels—byte, protocol, and packet—capturing comprehensive traffic structural information.
During fine-tuning, Nethira proposes a consistency-regularized strategy with hierarchical traffic augmentation to reduce label dependence.
Experiments on four public datasets demonstrate that Nethira outperforms seven existing pre-trained models, achieving an average F1-score improvement of 9.11\%, and reaching comparable performance with only 1\% labeled data on high-heterogeneity network tasks.

\end{abstract}


\section{Introduction}
\label{sec:Introduction}

Network traffic classification, which aims to organize network traffic into different categories such as applications and services, is fundamental and vital for network security and management \cite{ICASSP,NAS-ETC,FlowPrint,TFEGNN,etbert,YaTC,TrafficFormer,CN2025}.
Traditional rule-based methods \cite{FlowPrint,AppScanner} classify network traffic using manually defined rules by extracting basic attributes such as protocols and port numbers.
With the widespread adoption of encryption protocols and the use of dynamic ports, these methods have become increasingly ineffective.
To address these challenges, machine learning and deep learning techniques have emerged, offering the ability to automatically extract valuable features from raw traffic and substantially improving classification performance \cite{Fs-Net,EBSNN,TFEGNN}.
Nevertheless, their effectiveness is highly dependent on large-scale labeled data, which are often impractical to obtain in real-world network environments \cite{NetAugment}.

Recent advances \cite{NetMamba,NetConv,YaTC,PERT,NetGPT,etbert,TrafficFormer,TraGe} are leveraging pre-training technology to enable efficient network traffic classification with small-scale labeled data.
This approach typically involves two stages: \textit{pre-training} on large-scale unlabeled data for general traffic representations, and \textit{fine-tuning} on small-scale labeled data for specific network tasks.
Several studies \cite{PERT,etbert,NetGPT,YaTC,NetMamba} have demonstrated the potential of applying pre-training techniques for network traffic classification.
For instance, ET-BERT \cite{etbert} flattens network traffic into byte sequences analogous to text sentences, leveraging masked language modeling and next-sentence prediction to capture byte-level representations.
NetGPT \cite{NetGPT} also relies on flattened byte sequences, but uses language modeling during pre-training.
Subsequent works \cite{TraGe,TrafficFormer} regard ``\textit{flattened byte sequence}" as the de facto choice, and on this basis, enhance the model’s ability to further capture traffic byte characteristics.
For example, TraGe \cite{TraGe} introduces a header–payload differentiation pre-training task that exploits the distinct byte distributions of headers and payloads, while TrafficFormer \cite{TrafficFormer} proposes a protocol-field randomization strategy to help the model focus on key byte information during fine-tuning.


Although flattening network traffic into byte sequences enables pre-trained models to learn fundamental byte-level representations, this choice forces model inputs into a homogeneous form.
Such homogeneous input (i.e., flattened byte sequence) fails to effectively represent the inherent heterogeneity of network traffic—where discriminative features arise at different levels (e.g., byte, protocol, packet)—which is critical for network tasks.
For instance, in network attack detection, attackers may hide payloads through \textit{byte} obfuscation, whereas abnormal \textit{protocol} fields and excessive \textit{packet} fragmentation remain revealing of malicious activity \cite{ciciot2022}.
Despite this, existing pre-trained models have no specialized built-in mechanisms to characterize these hierarchical traffic structures (i.e., heterogeneity unawareness).

\begin{figure*}[t!]
\vspace{-5pt}
\centerline{\includegraphics[width=1 \linewidth]{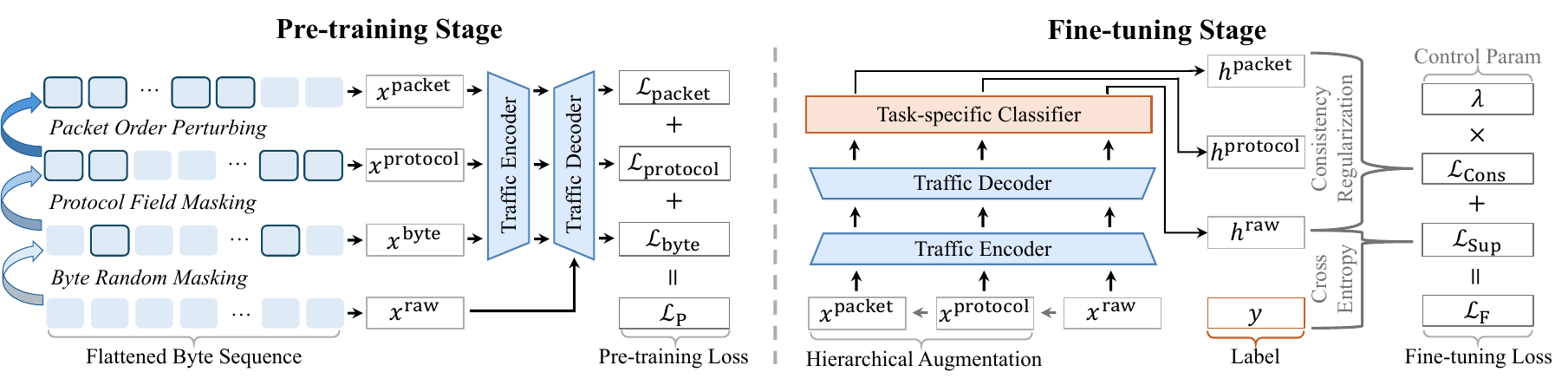}}
\vspace{-15pt}
\caption{The framework of Nethira.}
\vspace{-10pt}
\label{fig:framework}
\end{figure*}

In this paper, we aim to equip the model with awareness of hierarchical heterogeneity in network traffic under homogeneous inputs (i.e., flattened byte sequences), thereby improving classification performance and reducing reliance on labeled data.
Our contributions are summarized as follows:

\begin{itemize}
    \vspace{-8pt}
    \item We propose Nethira, a heterogeneity-aware hierarchical pre-trained model for network traffic classification. Nethira can form a general traffic representation by modeling its heterogeneity via hierarchical reconstruction and augmentation, thus reducing label dependence.
    \vspace{-20pt}
    \item We design a pre-training task based on hierarchical reconstruction that masks and perturbs bytes, protocols, and packets, explicitly guiding the model to be aware of hierarchical traffic heterogeneity and ultimately forming a general traffic representation.
    \vspace{-8pt}
    \item We introduce a fine-tuning strategy based on hierarchical augmentation that applies multi-level traffic data augmentation with consistency regularization, enhancing model generalization against heterogeneous traffic data and reducing reliance on labeled data.
    \vspace{-8pt}
    \item We evaluate Nethira against state-of-the-art methods on four public datasets. Nethira achieves an average F1-score improvement of 9.11\% compared to seven pre-training models, and reaches comparable performance with only 1\% labeled data on high-heterogeneity tasks. 
    \vspace{-15pt}
\end{itemize}

\vspace{-20pt}
\section{Method}
\vspace{-5pt}
\label{sec:Method}

\vspace{-5pt}
\subsection{Framework}
\vspace{-3pt}
\label{sec:Framework}

The overall framework of Nethira is illustrated in Fig. \ref{fig:framework}, and consists of two stages: pre-training and fine-tuning.
In the pre-training stage, network traffic is first transformed into a flattened byte sequence.
Then, we apply hierarchical perturbations at three levels:
\textit{byte} random masking, \textit{protocol} field masking, and \textit{packet} order perturbing.
Using a Transformer-based encoder–decoder architecture, the model is then pre-trained through byte sequence reconstruction.
To address the significant traffic heterogeneity during the fine-tuning stage, we employ a consistency regularization strategy based on hierarchical traffic data augmentation.
This strategy enhances the model’s ability to capture traffic heterogeneity while reducing its label dependence.


\vspace{-9pt}
\subsection{The Pre-training Stage}
\vspace{-2pt}
\label{sec:Pre-training}

During the pre-training stage, the model is trained on large-scale unlabeled traffic data.
The traffic data is first transformed into byte sequences, which serve as the input for pre-training. 
Then, the model is pre-trained through hierarchical sequence reconstruction, capturing traffic characteristics across multiple levels—byte, protocol, and packet.

\subsubsection{Traffic Data Preprocessing}
\vspace{-3pt}
\label{sec:Data Preprocessing}
Raw network traffic is first segmented into flows defined by the five-tuple (IPsrc:PORTsrc, IPdst:PORTdst, Protocol).
We preprocess each network flow to remove address bias and ensure length consistency.
Specifically, we set the Ethernet address, IP address, and TCP/UDP port fields to zero to eliminate address-related bias information.
Then, to manage variability in packet size within a flow, we select the first $M$ packets.
Each packet’s byte sequence is standardized to a fixed length $L$: sequences longer than $L$ are truncated, while shorter sequences are zero-padded.
The byte sequences of the first $M$ packets $p_{i}$ in the network flow are concatenated into a flattened byte sequence to construct the model input $x$:
{\setlength{\abovedisplayskip}{2pt}
 \setlength{\belowdisplayskip}{2pt}
\begin{equation}
x = \text{Concat}(p_{1}, p_{2}, \ldots, p_{M}) = (b_{1}, b_{2}, \ldots, b_{M \times L})
\end{equation}}

\noindent where $b_{i}$ denotes the $i$-th byte, with $M$ = 5 and $L$ = 128.

\vspace{-6pt}
\subsubsection{Model Pre-training}
\vspace{-3pt}
\label{sec:Pre-training Task}

We first present the model structure of Nethira, then introduce the three pre-training tasks for Nethira separately.

\noindent \textbf{Model Structure.}
Nethira adopts an encoder–decoder structure consisting of three components: byte embedding, traffic encoder, and traffic decoder. 
The byte embedding layer transforms raw bytes into high-dimensional vectors.
The traffic encoder, composed of stacked Transformer \cite{Transformer} encoders, captures global byte-level representations via self-attention.
The Transformer-based traffic decoder autoregressively reconstructs byte sequences to model contextual dependencies and enable pre-training through sequence reconstruction.

\noindent \textbf{Byte-level Reconstruction.}
Byte sequences in network traffic encode critical information like encryption.
Modeling these sequences enhances the ability to capture traffic patterns and improves generalization.
Therefore, we propose a byte-level reconstruction task: a set of byte positions $\mathcal{M}_{\text{byte}}$ is randomly sampled from the input sequence, the corresponding bytes $\tilde{b}^{\text{byte}}_t$ are replaced with a special token [MASK], and the model is pre-trained to reconstruct the original bytes.
{\setlength{\abovedisplayskip}{2pt}
 \setlength{\belowdisplayskip}{2pt}
 \begin{equation}
\mathcal{L}_{\text{byte}} = -
\sum_{t \in \mathcal{M}_{\text{byte}}}
\log P_\theta \left( b_t \mid \tilde{b}^{\text{byte}}_1, \ldots, \tilde{b}^{\text{byte}}_{M \times L} \right)
\end{equation}}

\noindent where $(\tilde{b}^{\text{byte}}_1, \ldots, \tilde{b}^{\text{byte}}_{M \times L})$ is the byte-level masked sequence, and $\theta$ is the model parameter of Nethira.


\noindent \textbf{Protocol-level Reconstruction.}
Network packet headers comprise multiple protocol fields (e.g., total length, TTL), represented as fixed-length consecutive byte sequences that encode transmission status and control information.
To model their sequential structure, we design a protocol-level reconstruction task that masks contiguous byte spans aligned with or contained within protocol fields.
Specifically, we mask $k$ consecutive bytes starting from positions restricted to field boundaries or their near vicinity, forming the masked byte position set $\mathcal{M}_{\text{protocol}}$.
This design ensures that most masked spans correspond to entire fields or substantial portions of them, rather than random byte fragments, thereby enhancing the model’s ability to learn general representations across different network protocols.
The model is pre-trained by reconstructing masked bytes $\tilde{b}^{\text{protocol}}_{t}$ from $\mathcal{M}_{\text{protocol}}$:
{\setlength{\abovedisplayskip}{3pt}
 \setlength{\belowdisplayskip}{2pt}
\begin{equation}
\mathcal{L}_{\text{protocol}} = - \sum_{t \in \mathcal{M}_{\text{protocol}}}
\log P_{\theta}\big(b_t \mid \tilde{b}^{\text{protocol}}_{1}, \ldots, 
\tilde{b}^{\text{protocol}}_{M \times L}\big)
\end{equation}}

\noindent \textbf{Packet-level Reconstruction.}
Packet sequences in network flows exhibit dynamic behaviors such as disorder and attack patterns.
Models relying solely on byte-level features fail to capture these cross-packet dependencies, necessitating explicit modeling of inter-packet structures.
We therefore propose a packet-level reconstruction task.
Specifically, the original packet order is perturbed to form a new packet sequence $(p_{\pi(1)}, \ldots, p_{\pi(M)})$, after which several byte positions are randomly masked to construct $\mathcal{M}_{\text{packet}}$.
The model is pre-trained to reconstruct the masked bytes $\tilde{b}^{\text{packet}}_{t}$, thereby learning packet-order variations and associated behavioral patterns. This process can
be formally expressed as follows:
{\setlength{\abovedisplayskip}{3pt}
 \setlength{\belowdisplayskip}{2pt}
\begin{equation}
(\bar{b}_1, \bar{b}_2, \ldots, \bar{b}_{M \times L}) 
= \text{Concat}(p_{\pi(1)}, \ldots, p_{\pi(M)})
\end{equation}}
\vspace{-7pt}
{\setlength{\abovedisplayskip}{2pt}
 \setlength{\belowdisplayskip}{0pt}
\begin{equation}
\mathcal{L}_{\text{packet}} = - \sum_{t \in \mathcal{M}_{\text{packet}}}
\log P_{\theta}\big(\bar{b}_t \mid \tilde{b}^{\text{packet}}_{1}, \ldots, 
\tilde{b}^{\text{packet}}_{M \times L}\big)
\end{equation}}

The pre-training objective $\mathcal{L}_{\text{P}}$ is defined as the sum of the three reconstruction losses above:
{\setlength{\abovedisplayskip}{2pt}
 \setlength{\belowdisplayskip}{2pt}
\begin{equation}
\mathcal{L}_{\text{P}} = \mathcal{L}_{\text{byte}} + \mathcal{L}_{\text{protocol}} + \mathcal{L}_{\text{packet}}
\end{equation}}

\vspace{-6pt}
\subsection{The Fine-tuning Stage}
\vspace{-3pt}
\label{sec:Fine-tuning}

During the fine-tuning stage, we introduce a consistency-regularized strategy based on hierarchical traffic data augmentation.
The objective is to improve Nethira's generalization
by enforcing prediction consistency under different levels of traffic heterogeneity augmentations.
Given an input byte sequence $x^{\text{raw}}$, two augmentation schemes are applied:

\begin{itemize}[leftmargin=*]
    \vspace{-6pt}
    \item \textbf{Protocol-level augmentation }$x^{\text{protocol}}$: We randomize the protocol field arrangement, preventing reliance on fixed orders and encouraging the model to learn full field distributions, thus improving generalization to unseen protocols.
    \vspace{-6pt}
     \item \textbf{Packet-level augmentation }$x^{\text{packet}}$: We perturb network packet sequences, simulating network dynamics such as reordering and partial loss. This training strategy enhances model generalization to dynamic network conditions.
    \vspace{-6pt}
\end{itemize}

Next, the three byte sequences $x^{\text{raw}}$, $x^{\text{protocol}}$, $x^{\text{packet}}$ are processed by the pre-trained traffic encoder, traffic decoder, and the task-specific classifier (a multi-layer MLP), yielding deep traffic representations $h^{\text{raw}}$, $h^{\text{protocol}}$, $h^{\text{packet}}$.

For labeled samples with ground-truth label $y$, the supervised loss is defined as the standard cross-entropy:
{\setlength{\abovedisplayskip}{2pt}
 \setlength{\belowdisplayskip}{2pt}
\begin{equation}
\mathcal{L}_{\text{sup}} = \mathrm{CE}(h^{\text{raw}}, y)
\end{equation}}

To enhance consistency under traffic heterogeneity, we further minimize the divergence between the deep traffic representations of the original input and its augmented variants. With Kullback–Leibler divergence as the metric:
{\setlength{\abovedisplayskip}{4pt}
 \setlength{\belowdisplayskip}{4pt}
\begin{equation}
\mathcal{L}_{\text{cons}} =  D_{KL}(h^
{\text{raw}}\,\|\, h^{\text{protocol}}) + D_{KL}(h^
{\text{raw}} \,\|\, h^{\text{packet}}) 
\end{equation}}

The fine-tuning loss $\mathcal{L}_{\text{F}}$ combines the two terms above:
{\setlength{\abovedisplayskip}{2pt}
 \setlength{\belowdisplayskip}{2pt}
\begin{equation}
\mathcal{L}_{\text{F}} = \mathcal{L}_{\text{sup}} + \lambda\,\mathcal{L}_{\text{cons}}
\end{equation}}

\noindent where $\lambda$ is a constant weight that balances the contribution of consistency regularization during the fine-tuning stage.

\vspace{-5pt}
\section{Evaluation}
\label{sec:Evaluation}

\begin{table*}[t!]
\renewcommand{\arraystretch}{0.6}
\vspace{-5pt}
\caption{Overall classification performance (\%) comparison of different methods on four network traffic datasets.}
\resizebox{0.98 \textwidth}{!}{%
\begin{tabular}{c|ccc|ccc|ccc|ccc|c}
\toprule
\multirow{2}{*}{Method} &
\multicolumn{3}{c|}{{ISCX-VPN(App)}} &
\multicolumn{3}{c|}{{ISCX-VPN(Service)}} &
\multicolumn{3}{c|}{{USTC-TFC}} &
\multicolumn{3}{c|}{{CIC-IoT}} &
\multirow{2}{*}{Avg. F1} \\

&
PR &
RC &
F1 &
PR &
RC &
F1 &
PR &
RC &
F1 &
PR &
RC &
F1 &
\\ \midrule

FlowPrint &
59.04 & 
43.04 & 
44.94 &
70.21 & 
66.62 & 
64.51 &
69.76 &
70.16 & 
68.81 &
14.73 &
20.46 &
15.70 &
48.49 \\

AppScanner &
72.89 &
53.61 &
58.03 &
85.99 &
75.67 &
79.13 &
75.58 &
57.72 &
62.77 &
35.27 &
23.86 &
25.45 &
56.35 \\



FS-Net &
49.90 &
39.96 &
40.60 &
71.61 &
63.63 &
64.18 &
90.74 &
89.66 &
89.39 &
37.24 &
35.39 &
32.61 &
56.70 \\

EBSNN &
66.07 &
61.53 &
62.05 &
89.84 &
89.69 &
89.53 &
93.48 &
91.29 &
90.10 &
88.92 &
87.29 &
85.37 &
81.76 \\

TFE-GNN &
67.20 &
60.60 &
61.80 &
85.97 &
80.95 &
82.14 &
95.91 &
95.68 &
95.63 &
67.05 &
66.90 &
64.29 &
75.97 \\ \midrule

NetMamba &
67.17 &
58.05 &
60.32 &
86.01 &
78.31 &
80.27 &
95.85 &
94.90 &
94.83 &
68.18 &
70.39 &
67.55 &
75.74 \\

YaTC &
70.03 &
58.73 &
62.33 &
81.06 &
78.37 &
78.06 &
95.77 &
94.96 &
94.87 &
74.28 &
75.07 &
72.36 &
76.91 \\

PERT &
72.16 &
70.26 &
70.80 &
91.42 &
90.43 &
90.86 &
93.24 &
93.00 &
92.95 &
89.58 &
89.47 &
88.23 &
85.71 \\

NetGPT &
69.86 &
71.48 &
69.40 &
91.94 &
92.20 &
91.92 &
96.16 &
95.98 &
96.00 &
90.48 &
90.19 &
89.08 &
86.60 \\

ET-BERT  &
72.00 &
70.36 &
70.94 &
91.40 &
91.58 &
91.47 &
95.21 &
95.20 &
95.18 &
91.29 &
89.93 &
88.91 &
86.63 \\

TraGe &
71.38 &
71.10 &
70.93 &
91.75 &
91.72 &
91.68 &
95.94 &
95.90 &
95.91 &
89.02 &
90.04 &
88.61 &
86.78 \\

TrafficFormer &
72.32 &
71.56 &
71.69 &
92.15 &
91.94 &
91.97 &
95.17 &
94.98 &
95.01 &
91.25 &
90.10 &
89.12 &
86.95 \\ \midrule

\textbf{Nethira} &
\textbf{77.33} &
\textbf{74.58} &
\textbf{75.55} &
\textbf{92.35} &
\textbf{92.44} &
\textbf{92.34} &
\textbf{96.62} &
\textbf{96.42} &
\textbf{96.40} &
\textbf{97.26} &
\textbf{97.40} &
\textbf{97.29} &
\textbf{90.40} \\

\bottomrule

\end{tabular}%
}
\label{tab:overall}
\end{table*}


\begin{figure*}[t!]
\vspace{-10pt}
\centerline{\includegraphics[width=1 \linewidth]{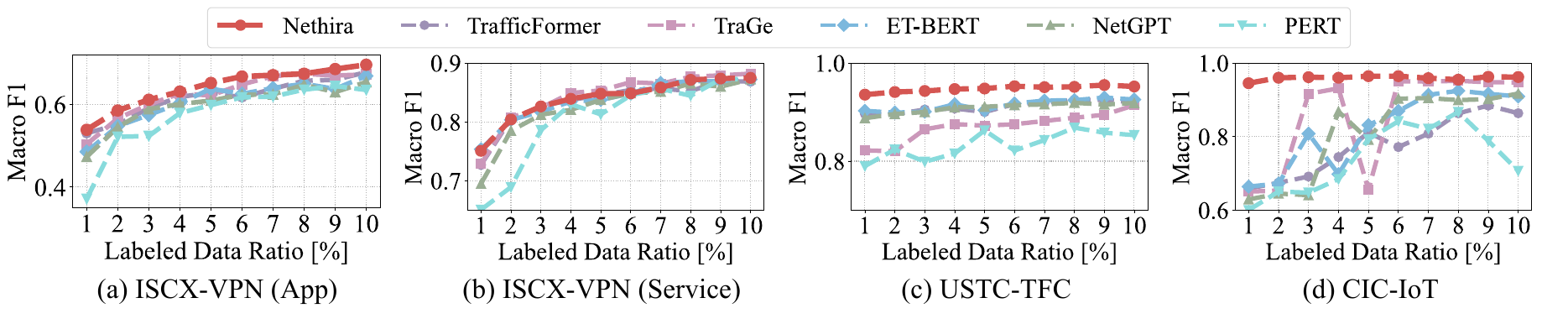}}
\vspace{-15pt}
\caption{Comparison results of classification performance on four network traffic datasets with limited labeled data.}
\vspace{-10pt}
\label{fig:infer}
\end{figure*}

\vspace{-5pt}
\subsection{Experimental Settings}
\label{sec:Experimental Settings}

\vspace{-5pt}
\noindent \textbf{Pre-training and Fine-tuning Dataset.}
To ensure a fair comparison among pre-trained models, we adopt the open-source corpus from ET-BERT \cite{etbert} for pre-training.
For downstream evaluation, we employ four datasets: ISCX-VPN (App) \cite{ISCX-VPN}, ISCX-VPN (Service) \cite{ISCX-VPN}, USTC-TFC \cite{USTC-TFC}, and CIC-IoT \cite{ciciot2022}.
The ISCX-VPN \cite{ISCX-VPN} dataset captures encrypted traffic from various applications and services, and we further split it into ISCX-VPN (App) and ISCX-VPN (Service) for application and service identification, respectively.
The USTC-TFC \cite{USTC-TFC} dataset, comprising 10 types of malicious traffic mainly from smart grid infrastructures, is used for malware classification.
The CIC-IoT \cite{ciciot2022} dataset, designed for attack identification in IoT environments, exhibits highly heterogeneous traffic patterns, as attackers often exploit diverse devices and adopt varied attack strategies.
Following existing pre-trained models \cite{etbert,NetGPT}, we randomly select at most 5000 flows per class, and split each dataset into training, validation, and testing sets in an 8:1:1 ratio.

\noindent \textbf{Implementation Details and Baselines.}
We conduct experiments using Python 3.10 and CUDA 12.4.
Nethira consists of six encoder and six decoder layers.
We pre-train Nethira for 100,000 steps with a learning rate of 1e-4.
During fine-tuning, we run 10 epochs, and set parameter $\lambda$ to 0.1.
We use three typical metrics: Precision (PR), Recall (RC), and F1-score (F1).
We select twelve baselines, including (1) two statistics features based models, i.e., FlowPrint \cite{FlowPrint} and AppScanner \cite{AppScanner}; (2) three deep learning based models, i.e., FSNet \cite{Fs-Net}, EBSNN \cite{EBSNN}, TFE-GNN \cite{TFEGNN}; and (3) seven pre-trained models, i.e., NetMamba \cite{NetMamba}, YaTC \cite{YaTC}, PERT \cite{PERT}, NetGPT \cite{NetGPT}, ET-BERT \cite{etbert}, TraGe \cite{TraGe}, and TrafficFormer \cite{TrafficFormer}.

\vspace{-7pt}
\subsection{Overall Classification Performance}
\vspace{-2pt}

Table \ref{tab:overall} presents the classification performance of Nethira compared to all baselines across four network traffic datasets.
Nethira consistently outperforms all baselines, achieving an average F1-score improvement of 9.11\% over seven pre-trained models.
Specifically, Nethira achieves improvements of 11.49\%, 5.36\%, 1.52\%, and 18.05\% on ISCX-VPN (App), ISCX-VPN (Service), USTC-TFC, and CIC-IoT, respectively.
Overall, the results demonstrate that Nethira surpasses all baselines by a substantial margin.
Specifically, Nethira's superior performance stems from its ability to learn general representations of heterogeneous traffic through hierarchical reconstruction during pre-training and hierarchical augmentation during fine-tuning.
In comparison, existing pre-trained models that emphasize byte-level features show slight improvements.
For instance, TrafficFormer reports an average F1-score improvement of 0.37\% compared to ET-BERT.

\vspace{-9pt}
\subsection{Performance Under Limited Labeled Data}
\vspace{-2pt}

We evaluate Nethira under limited labeled data by reducing the labeled training set to 1–10\% of its original size, as shown in Fig. \ref{fig:infer}. 
The results show that Nethira significantly reduces reliance on labeled data compared to baseline methods.
For example, on the CIC-IoT dataset, Nethira achieves an F1-score of 0.9452 with only 1\% labeled data, outperforming all pre-trained models and even surpassing several models trained with 100\% labels, such as TrafficFormer (0.8912 in Table \ref{tab:overall}).
In addition, on ISCX-VPN (App) under the same limited-label setting, Nethira delivers performance comparable to existing pre-trained models rather than significant improvements.
We attribute this cross-dataset difference to traffic heterogeneity.
Using the average number of packets per flow (ANPF) as a metric, the ANPFs of the CIC-IoT and ISCX-VPN (App) datasets are 12 and 2, respectively.
A higher ANPF increases protocol diversity and amplifies the influence of network dynamics on packet order—conditions under which Nethira is particularly effective.
Overall, these results demonstrate that with just 1\% labeled data, Nethira can match or even surpass models trained with 100\% labeled data, highlighting its reduced dependence on labeled data.

\vspace{-5pt}
\subsection{Ablation Study}
\vspace{-2pt}

We conduct ablation studies on Nethira using the CIC-IoT dataset. 
First, removing the pre-training stage and training the model from scratch results in a 4.78\% performance drop, underscoring its role in learning general traffic representations.
Second, substituting our hierarchical reconstruction-based pre-training with a basic byte-mask prediction task ($\mathcal{L}_{\text{byte}}$ only) reduces performance by 1.71\%, showing that hierarchical reconstruction better captures traffic heterogeneity and improves model performance.
Third, replacing our hierarchical augmentation-based fine-tuning with standard supervised fine-tuning ($\mathcal{L}_{\text{sup}}$ only) causes a 7.84\% performance decline, demonstrating that our fine-tuning strategy significantly enhances model learning from heterogeneous traffic, thereby improving downstream classification performance.

\vspace{-5pt}
\section{Conclusion}
\label{sec:Conclusion}
\vspace{-5pt}

In this paper, we propose Nethira, a heterogeneity-aware hierarchical pre-trained model for network traffic classification.
Nethira consists of hierarchical reconstruction-based pre-training and hierarchical augmentation-based fine-tuning.
Experiments on four datasets show Nethira outperforms state-of-the-art baselines and reduces reliance on labeled data.

\vfill\pagebreak

\clearpage

\section{Acknowledgment}
\vspace{-5pt}

We thank all the anonymous reviewers. This work was supported in whole or in part, by the Strategic Priority Research Program of Chinese Academy of Sciences under Grant No.XDB0500103, by the National Natural Science Foundation of China under Grant Nos. U24B6012, U2333201, and 62372429, and in part by the Innovation Funding of ICT, CAS under Grant No. E461040.
\bibliographystyle{IEEEbib}
\vspace{-10pt}
\bibliography{reference}

@inproceedings{etbert,
  title={Et-bert: A contextualized datagram representation with pre-training transformers for encrypted traffic classification},
  author={Lin, Xinjie and Xiong, Gang and Gou, Gaopeng and Li, Zhen and Shi, Junzheng and Yu, Jing},
  booktitle={Proceedings of the ACM Web Conference},
  pages={633--642},
  year={2022}
}

@article{NetGPT,
  title={NetGPT: Generative Pretrained Transformer for Network Traffic},
  author={Meng, Xuying and Lin, Chungang and Wang, Yequan and Zhang, Yujun},
  journal={arXiv preprint arXiv:2304.09513},
  year={2023}
}

@inproceedings{TFEGNN,
  title={Tfe-gnn: A temporal fusion encoder using graph neural networks for fine-grained encrypted traffic classification},
  author={Zhang, Haozhen and Yu, Le and Xiao, Xi and Li, Qing and Mercaldo, Francesco and Luo, Xiapu and Liu, Qixu},
  booktitle={Proceedings of the ACM Web Conference},
  year={2023}
}

@inproceedings{FlowPrint,
  title={Flowprint: Semi-supervised mobile-app fingerprinting on encrypted network traffic},
  author={Van Ede, Thijs and Bortolameotti, Riccardo and Continella, Andrea and Ren, Jingjing and Dubois, Daniel J and Lindorfer, Martina and Choffnes, David and van Steen, Maarten and Peter, Andreas},
  booktitle={Network and distributed system security symposium},
  volume={27},
  year={2020}
}

@inproceedings{AppScanner,
  title={Appscanner: Automatic fingerprinting of smartphone apps from encrypted network traffic},
  author={Taylor, Vincent F and Spolaor, Riccardo and Conti, Mauro and Martinovic, Ivan},
  booktitle={2016 IEEE European Symposium on Security and Privacy (EuroS\&P)},
  pages={439--454},
  year={2016},
  organization={IEEE}
}

@inproceedings{Fs-Net,
  title={Fs-net: A flow sequence network for encrypted traffic classification},
  author={Liu, Chang and He, Longtao and Xiong, Gang and Cao, Zigang and Li, Zhen},
  booktitle={IEEE INFOCOM 2019-IEEE Conference On Computer Communications},
  pages={1171--1179},
  year={2019},
  organization={IEEE}
}

@article{EBSNN,
  title={EBSNN: Extended byte segment neural network for network traffic classification},
  author={Xiao, Xi and Xiao, Wentao and Li, Rui and Luo, Xiapu and Zheng, Haitao and Xia, Shutao},
  journal={IEEE Transactions on Dependable and Secure Computing},
  volume={19},
  number={5},
  pages={3521--3538},
  year={2021},
  publisher={IEEE}
}

@inproceedings{YaTC,
  title={Yet another traffic classifier: A masked autoencoder based traffic transformer with multi-level flow representation},
  author={Zhao, Ruijie and Zhan, Mingwei and Deng, Xianwen and Wang, Yanhao and Wang, Yijun and Gui, Guan and Xue, Zhi},
  booktitle={Proceedings of the AAAI Conference on Artificial Intelligence},
  volume={37},
  number={4},
  pages={5420--5427},
  year={2023}
}

@inproceedings{PERT,
  title={PERT: Payload encoding representation from transformer for encrypted traffic classification},
  author={He, Hong Ye and Yang, Zhi Guo and Chen, Xiang Ning},
  booktitle={ITU K},
  pages={1--8},
  year={2020},
  organization={IEEE}
}

@inproceedings{ISCX-VPN,
  title={Characterization of encrypted and vpn traffic using time-related},
  author={Draper-Gil, Gerard and Lashkari, Arash Habibi and Mamun, Mohammad Saiful Islam and Ghorbani, Ali A},
  booktitle={ICISSP},
  pages={407--414},
  year={2016}
}

@inproceedings{USTC-TFC,
  title={Malware traffic classification using convolutional neural network for representation learning},
  author={Wang, Wei and Zhu, Ming and Zeng, Xuewen and Ye, Xiaozhou and Sheng, Yiqiang},
  booktitle={International conference on information networking},
  pages={712--717},
  year={2017},
  organization={IEEE}
}

@inproceedings{ciciot2022,
  title={Towards the development of a realistic multidimensional IoT profiling dataset},
  author={Dadkhah, Sajjad and Mahdikhani, Hassan and Danso, Priscilla Kyei and Zohourian, Alireza and Truong, Kevin Anh and Ghorbani, Ali A},
  booktitle={PST},
  pages={1--11},
  year={2022},
  organization={IEEE}
}

@inproceedings{NetAugment,
  title={Realistic Website Fingerprinting By Augmenting Network Traces},
  author={Bahramali, Alireza and Bozorgi, Ardavan and Houmansadr, Amir},
  booktitle={Proceedings of the 2023 ACM SIGSAC Conference on Computer and Communications Security},
  pages={1035--1049},
  year={2023}
}

@inproceedings{TrafficFormer,
  title={Trafficformer: an efficient pre-trained model for traffic data},
  author={Zhou, Guangmeng and Guo, Xiongwen and Liu, Zhuotao and Li, Tong and Li, Qi and Xu, Ke},
  booktitle={2025 IEEE Symposium on Security and Privacy},
  pages={1844--1860},
  year={2025},
  organization={IEEE}
}

@inproceedings{NetMamba,
  title={Netmamba: Efficient network traffic classification via pre-training unidirectional mamba},
  author={Wang, Tongze and Xie, Xiaohui and Wang, Wenduo and Wang, Chuyi and Zhao, Youjian and Cui, Yong},
  booktitle={32nd International Conference on Network Protocols},
  pages={1--11},
  year={2024},
  organization={IEEE}
}

@inproceedings{TraGe,
  title={TraGe: A Generic Packet Representation for Traffic Classification Based on Header-Payload Differences},
  author={Lin, Chungang and Jiang, Yilong and Zhang, Weiyao and Meng, Xuying and Zuo, Tianyu and Zhang, Yujun},
  booktitle={2025 IEEE/ACM 29th International Symposium on Quality of Service (IWQOS)},
  pages={1--6},
  year={2025},
  organization={IEEE}
}

@article{Transformer,
  title={Attention is all you need},
  author={Vaswani, Ashish and Shazeer, Noam and Parmar, Niki and Uszkoreit, Jakob and Jones, Llion and Gomez, Aidan N and Kaiser, {\L}ukasz and Polosukhin, Illia},
  journal={NeurIPS},
  volume={30},
  year={2017}
}

@inproceedings{NAS-ETC,
  title={ANASETC: Automatic Neural Architecture Search for Encrypted Traffic Classification},
  author={Zhang, Heng and Chen, Ziqian and Xia, Wei and Xiong, Gang and Gou, Gaopeng and Li, Zhen and Huang, Guangyan and Li, Yunpeng},
  booktitle={ICASSP 2025-2025 IEEE International Conference on Acoustics, Speech and Signal Processing (ICASSP)},
  pages={1--5},
  year={2025},
  organization={IEEE}
}

@inproceedings{ICASSP,
  title={A multi-modal approach for context-aware network traffic classification},
  author={Pang, Bo and Fu, Yongquan and Ren, Siyuan and Shen, Siqi and Wang, Ye and Liao, Qing and Jia, Yan},
  booktitle={ICASSP 2023-2023 IEEE International Conference on Acoustics, Speech and Signal Processing (ICASSP)},
  pages={1--5},
  year={2023},
  organization={IEEE}
}

@article{NetConv,
  title={Convolutions are Competitive with Transformers for Encrypted Traffic Classification with Pre-training},
  author={Lin, Chungang and Zhang, Weiyao and Zuo, Tianyu and Zha, Chao and Jiang, Yilong and Meng, Ruiqi and Luo, Haitong and Meng, Xuying and Zhang, Yujun},
  journal={arXiv preprint arXiv:2508.02001},
  year={2025}
}

@article{CN2025,
  title={Accelerating traffic engineering optimization for segment routing: A recommendation perspective},
  author={Wang, Linghao and Wang, Miao and Lin, Chungang and Zhang, Yujun},
  journal={Computer Networks},
  volume={264},
  pages={111224},
  year={2025},
  publisher={Elsevier}
}

\end{document}